\documentclass[runningheads]{llncs}
\usepackage{graphicx}
\usepackage{bm}
\usepackage{amsmath}
\usepackage{amsfonts}
\usepackage{hyperref}
\usepackage{graphics}
\usepackage{tabularx}
\usepackage{booktabs}
\usepackage{color}
\usepackage{multirow}
\usepackage{siunitx}
\usepackage{appendix}

\begin{document}
\title{Relaxometry Guided Quantitative Cardiac Magnetic Resonance Image Reconstruction}
\titlerunning{Relaxometry guided QMRI Reconstruction}
%
\author{Yidong Zhao \inst{1}
\and Yi Zhang\inst{1} 
\and Qian Tao\inst{1}
}
%
\authorrunning{Y. Zhao et al.}
\institute{Department of Imaging Physics, Delft University of Technology, The Netherlands
\email{\{y.zhao-8, y.zhang-43, q.tao\}@tudelft.nl}}
\maketitle              
\begin{abstract}
Deep learning-based methods have achieved prestigious performance for magnetic resonance imaging (MRI) reconstruction, enabling fast imaging for many clinical applications. Previous methods employ convolutional networks to learn the image prior as the regularization term. In quantitative MRI, the physical model of nuclear magnetic resonance relaxometry is known, providing additional prior knowledge for image reconstruction. However, traditional reconstruction networks are limited to learning the spatial domain prior knowledge, ignoring the relaxometry prior. Therefore, we propose a relaxometry-guided quantitative MRI reconstruction framework to learn the spatial prior from data and the relaxometry prior from MRI physics. Additionally, we also evaluated the performance of two popular reconstruction backbones, namely, recurrent variational networks (RVN) and variational networks (VN) with U-Net. Experiments demonstrate that the proposed method achieves highly promising results in quantitative MRI reconstruction. Our code can be found at \url{https://github.com/pandafriedlich/relax_qmri_recon.git}.
\keywords{Caridac MRI  \and Quantitative mapping \and Relaxometry \and Image reconstruction.}
\end{abstract}
\section{Introduction}
Quantitative Magnetic Resonance Image (qMRI) has emerged as an indispensable imaging modality in research and clinical applications thanks to its quantitative measurements of tissue properties, such as $T_1$ and $T_2$ relaxation times \cite{seraphim2020quantitative}. A common approach for qMRI typically involves a two-step process. Firstly, multiple k-space data of a subject are acquired and reconstructed into a series of weighted images with varying imaging parameters (such as echo times and diffusion weights). Subsequently, the underlying tissue properties are estimated by fitting a signal model to the images \cite{shafieizargar2023systematic}. However, acquiring fully sampled k-space data, adhering to the Nyquist criterion, for multiple measurements in qMRI requires a time-consuming endeavor \cite{haacke1999magnetic}. This extended acquisition process introduces the potential for motion artifacts and leads to patient discomfort due to prolonged scan duration.

Motivated by reducing the acquisition time in qMRI while keeping the reconstruction and estimation quality, \textit{accelarated qMRI} has become one of the central research 
topics in qMRI. The methods in accelerated qMRI can be divided into two categories according to whether the method includes an intermediate step of image reconstruction from k-space data: If so, the method is categorized as \textit{indirect reconstruction}; Otherwise, the method is categorized as \textit{direct reconstruction} where only the parameter mapping estimations are gained \cite{shafieizargar2023systematic}. This paper will focus on indirect reconstruction methods since the goals require both reconstructing images and estimating the mapping parameters. However, our proposed method is flexible yet novel, taking self-supervised quantitative mapping as auxiliary constraints to guide the reconstruction.

In conventional qMRI, the acceleration can be achieved by parallel imaging (PI) \cite{heidemann2003brief,larkman2007parallel} or compressed sensing (CS) \cite{donoho2006compressed}. With specific undersampling patterns, PI utilizes multiple radio-frequency (RF) coils simultaneously with individual coil sensitivity maps to reduce the aliasing artifacts. The applications of PI include SENSE \cite{pruessmann2001SENSE}, which is the original implementation, and GRAPPA \cite{griswold2002GRAPPA}, which allows auto-calibration when sensitivity maps are missing. Meanwhile, CS-based methods achieve acceleration by using partial sampling in k-space, which loosens the requirement of the Nyquist criterion with sparsity assumption in true signals. Though the CS-based methods are well-established, they still struggle with the design of regularizer functions, which varies across different tissues \cite{lustig2007sparse}. 

Accelerated qMRI reconstruction benefits from recent rapid development in deep learning on both computer vision and inference learning since MRI reconstruction can be formulated as an inverse problem: Given a (partial) measurement in k-space, the target is to recover the image signal as close as possible. Deep learning-based methods for MRI reconstruction can be categorized into \textit{iterative} methods \cite{yang2018admm,hammernik2018learning,lonning2019recurrent,sriram2020end,yiasemis2022recurrent} and \textit{one-step} methods \cite{zhu2018image,hyun2018deep}. One-step methods aim to predict the refined reconstructed images from images reconstructed from corrupted 
 k-space \cite{hyun2018deep} or partially sampled k-space measurements directly \cite{zhu2018image}. Iterative methods leverage neural networks to learn an incremental refining process, analogous to a conventional optimization process. As a first attempt, Yang et al. \cite{yang2018admm} proposed ADMM-Net: a neural network to parameterize the alternating direction method of multipliers (ADMM), thus solving the inverse problem of MRI reconstruction in combination with CS. Similarly, Hammernik et al. proposed a variational network \cite{hammernik2018learning} in an unrolled gradient descent scheme for generalized CS reconstruction. Inspired by a more general idea in meta-learning, recurrent inference machines (RIM) \cite{putzky2017recurrent} were designed to find a maximum-a-posteriori (MAP) estimation to solve inverse problems when the corresponding forward model is known. This is then applied in MRI reconstruction \cite{lonning2019recurrent}, where in the RIM framework, the parameters are shared across the iterations with internal hidden states instead of individual units for each iteration in previous works. A more recent work, RecurrentVarNet \cite{yiasemis2022recurrent}, combines variational networks and RIM with a hybrid domain-learning strategy using convolutional neural networks (CNN) to guide the optimization in k-space. 

Unlike naive MRI reconstruction, in qMRI reconstruction, the involvement of physical signal models in methodological design can further improve the quality of both reconstruction and parameter mapping estimation since an anatomical correspondence across images is often assumed when designing the acquisition pipeline \cite{huizinga2016pca}. Apart from conventional least-square methods, parameter mappings can be estimated by dictionary matching~\cite{haaf2017cardiac,o2022t2,shafieizargar2023systematic}. Given the reconstructed images or raw k-space measurements, there are several methods utilizing self-supervised learning to estimate parameter mappings~\cite{liu2019mantis,barbieri2021deep,liu2021magnetic}. However, few existing works using ~\cite{eliasi2014fast} address the joint tasks of reconstruction and parameter estimation, while none lie in the deep learning paradigm.
\subsection{Contributions}
In this work, we tackle the quantitative mapping problem in the CMRxRecon challenge and make the following contributions: 
\begin{itemize} 
    \item We introduce a novel quantitative mapping network that learns to mimic MR physics in an unsupervised fashion. 
    \item We evaluate two different CNN architectures for image prior learning in the variational reconstruction network.
    \item We leverage the quantitative mapping network to guide the reconstruction process, ensuring that the output signal conforms to the MR relaxometry. 
\end{itemize}
\section{Methods}
\subsection{Parallel Imaging}
\subsubsection{Accelerated MRI} In this paper, we focus on the reconstruction of $k_t$ qMRI baseline images $\bm{x}\in \mathbb{C}^{k_x \times k_y \times k_t}$\footnote{We use $\bm{x}$ for both complex and magnitude image for simplicity.}, given their corresponding k-space measurements $\bm{y} \in \mathbb{C}^{n_c \times k_x \times k_y \times k_t}$ of $n_c$ coils, where $k_x$ and $k_y$ are image shapes along the frequency and phase encoding axes, respectively. The $c^{\mathrm{th}}$ coil measurement $\bm{y}_c$ is formulated as 
\begin{equation}
    \bm{y}_c = U\mathcal{F} S_c \bm{x} + \bm{\epsilon}_c,
\end{equation}
where $\mathcal{F}$ is the 2D Fourier transform operator, $S_c$ denotes the sensitivity map, $U$ characterises the Cartesian under-sampling pattern and $\bm{\epsilon}_c$ represents the measurement noise. The overall forward operator and its adjoint operator in accelerated parallel imaging can be formulated as $\mathcal{A} = U \circ \mathcal{F} \circ \mathcal{E} $ and $\mathcal{A}^{*} = \mathcal{R} \circ \mathcal{F} \circ U$~\cite{yiasemis2022recurrent}, where 
\begin{align}
    \mathcal{E} (\bm{x}) &= \left[S_1 \bm{x}, S_2 \bm{x}, \cdots, S_{n_c} \bm{x}\right], \\
    \mathcal{R}(\bm{y}) &= \sum_{c=1}^{n_c}S_c^{*} \bm{y}_c.
\end{align}
\subsubsection{Sensitivity Estimation} The sensitivity maps are estimated using the auto-calibration region $U_{\mathrm{AC}}$ in k-space which is always sampled in the low-frequency band. The initial sensitivity maps can be estimated as
\begin{equation}
    \hat{S}_c^{0} = \frac{\mathcal{F}^{-1}U_{\mathrm{AC}} \bm{y}_c}{\mathrm{RSS}\left (\left\{ \mathcal{F}^{-1}U_{AC} \bm{y}_l \right\}_{l=1}^{n_c}\right )},
    \label{eq:sense}
\end{equation}
where $\mathrm{RSS}$ denotes the root-sum-of-square operator defined in~\cite{larsson2003snr}
\begin{equation}
    \mathrm{RSS}(\bm{x}_1, \bm{x}_2, \cdots, \bm{x}_{n_c}) = \sqrt{\sum_{c=1}^{n_c} \vert \bm{x}_c \vert^2}.
\end{equation}
\begin{figure}[t]
    \centering
    \includegraphics{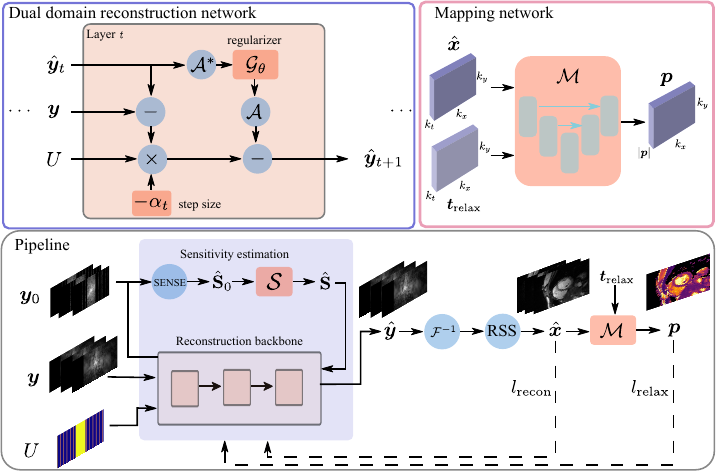}
    \caption{The reconstruction backbone consists of unrolled gradient descent layers, and the image prior is learned during training by $\mathcal{G}_{\theta}$. A pre-trained mapping network $\mathcal{M}$ is introduced to predict the quantitative parameters $\bm{p}$ and guide the reconstruction with MR relaxometry.}
    \label{fig:method_overview}
\end{figure}
\subsection{Relaxometry Guided Reconstruction}
\subsubsection{Dual-domain Reconstruction Network}
Reconstruction of an image $\bm{x}$ given its under-sampled parallel imaging measurements $\bm{y}$ can be formulated as optimizing the Lagrangian $\mathcal{L}(\bm{x})$ defined as
\begin{equation}
    \underset{\hat{\bm{x}}}{\operatorname{\arg \min}} \  \mathcal{L} (\bm{x}) = \Vert \mathcal{A} \hat{\bm{x}} - \bm{y} \Vert_2^2 + R(\hat{\bm{x}}),
\label{eq:lagrangian}
\end{equation} 
where the first term constrains the data fidelity via a $2-norm$ operator and the latter term $R(\cdot)$ is a regularizer to stabilize the solution space. The update rule of gradient-descent for optimization of Eqn.~\ref{eq:lagrangian} is 
\begin{align}
    \hat{\bm{x}}_{t+1} &= \hat{\bm{x}}_t -\alpha_{t}\frac{\partial \mathcal{L}}{\partial \hat{\bm{x}}^{*}}  \nonumber \\
            &= \hat{\bm{x}}_t - \alpha_t \mathcal{A}^{*}\left (\mathcal{A} \hat{\bm{x}}_t - \bm{y} \right ) - \mathcal{G}_{\theta_t} (\hat{\bm{x}}_t),
    \label{eq:update_image_domain}
\end{align}
where $\mathcal{G}_{\theta_t}$ is a convolutional network that learns the scaled Wirtinger derivatives $\frac{\partial R}{\partial \hat{\bm{x}}_t^*}$ of the regularizer \textit{w.r.t.} the current reconstruction, and $\alpha_t$ is a trainable step scalar. Applying Fourier transform on both sides of Eqn.~\ref{eq:update_image_domain} yields
\begin{align}
    \hat{\bm{y}}_{t+1} = \hat{\bm{y}}_t - \alpha_t U (\hat{\bm{y}}_{t} - \bm{y}) - \mathcal{F} \circ \mathcal{E} \circ \mathcal{G}_{\theta_t} (\mathcal{R} \circ \mathcal{F}^{-1} \hat{\bm{y}}_t),
    \label{eq:update_kspace}
\end{align}
in which the data fidelity part is updated in k-space and the regularization is performed in the spatial domain. 
In this work, we follow~\cite{hammernik2018learning,yiasemis2022recurrent} and use the variational network based on the unrolled update rule defined in Eqn.~\ref{eq:update_kspace} for image construction. Specifically, we investigate and compare two different types of CNNs to learn the regularizer $\mathcal{G}_{\theta_t}$: U-Net~\cite{ronneberger2015u} and convolutional Gated Recurrent Unit (GRU) blocks as in~\cite{yiasemis2022recurrent}. Additionally, we employ another U-Net $\mathcal{S}$ to refine the initial sensitivity map estimated in Eqn.~\ref{eq:sense} $\hat{S}_c = \mathcal{S}(\hat{S}_c^{0})$. The reconstruction network is trained with a combination of $L_1$ loss and structural similarity index measure (SSIM)~\cite{wang2004image}:
\begin{align}
     l_{\mathrm{recon}}(\hat{\bm{x}}) = \gamma_1 \Vert \hat{\bm{x}} - \bm{x} \Vert_1 + \gamma_2 \mathrm{SSIM}(\hat{\bm{x}}, \bm{x}),
\end{align}
where $\hat{\bm{x}}$ and $\bm{x}$ denote the predicted image and the fully sampled image, respectively. The architecture of reconstruction backbone layers is shown on the upper left panel in Fig.~\ref{fig:method_overview}.

\subsubsection{Quantitative Mapping Network} Given reconstructed magnitude image $\bm{x}\in \mathbb{R}^{k_x \times k_y \times k_t}$, quantitative mapping is usually treated as a parameter fitting problem and solved by least squares or dictionary matching~\cite{haaf2017cardiac,o2022t2,shafieizargar2023systematic}. Voxels in a reconstructed image of good quality should conform to the MR relaxometry and thus have a relatively lower level of fitting error since tissue anatomies are assumed to be spatially aligned. Inspired by this, we propose using the parameter fitting error to guide the reconstruction procedure. However, both least squares and dictionary matching are not differentiable and thus cannot be integrated into the computational graph. To make the mapping procedure differentiable, we propose using a U-Net $\mathcal{M}$ to predict the parameters given the image $\bm{x}$ and the inversion time or echo time $\bm{t}_{\mathrm{relax}}$. With the MR relaxation physics known, the network $\mathcal{M}$ is trained in a purely unsupervised fashion, and the physics-informed training loss $l_{\mathrm{relax}}$ is defined as:
\begin{align}
    l_{\mathrm{relax}}(\bm{x}) = \left\Vert s\left(\mathcal{M}\left(\bm{x}, \bm{t}_{\mathrm{relax}}\right), \bm{t}_{\mathrm{relax}}\right) - \bm{x} \right\Vert_1, 
\end{align}
where $s(\cdot)$ can be either $s_{T_1}$ or $s_{T_2}$ which characterises the signal intensity models for $T_1$ or $T_2$ relaxation:
\begin{align}
    s_{T_1}(A, B, T_1^*, t_{\mathrm{relax}}) &= \left \vert A -  B \exp\left(-\frac{t_{\mathrm{relax}}}{T_1^*}\right) \right \vert, \label{eq:t1} \\
    s_{T_2} (A, T_2, t_{\mathrm{relax}}) &= A  \exp \left( -\frac{t_{\mathrm{relax}}}{T_2}\right). \label{eq:t2}
\end{align}
Note that from Eqn.~\ref{eq:t1}, the parameter of interest $T_1$ is derived by $T_1 = (B/A-1)T_1^*$.
\subsubsection{Joint Reconstruction and Quantitative Mapping} The mapping network $\mathcal{M}$ is pre-trained with the fully sampled images only and is frozen during the reconstruction network training process. The physics-informed loss $l_{\mathrm{relax}}$ can then be used to enforce a relaxation physics-informed reconstruction. We also penalize the difference between the mapping prediction on fully sampled image $\mathcal{M}(\bm{x})$ and on the predicted image $\mathcal{M}(\hat{\bm{x}})$, such that the reconstructed image has a consistent parameter map as the fully sampled image. The total training loss is then formulated as:
\begin{equation}
    l(\hat{\bm{x}}) = l_{\mathrm{recon}}(\hat{\bm{x}}) + \gamma_3 l_{\mathrm{relax}}(\hat{\bm{x}}) + \gamma_4 \Vert\mathcal{M} (\hat{\bm{x}}) - \mathcal{M} (\bm{x})\Vert_1.
    \label{eq:total_loss}
\end{equation}
An overview of the proposed method is illustrated in Fig.~\ref{fig:method_overview}.

\section{Experiments}
\subsection{Dataset}
We conduct the experiments on the CMRxRecon challenge data\footnote{\href{https://cmrxrecon.github.io/}{https://cmrxrecon.github.io/}}~\cite{wang2023cmrxrecon}, consisting of 120 subjects for training. Imaging was performed on a Siemens 3T MRI scanner (MAGNETOM Vida) and the multi-coil images were compressed to 10 virtual coils, with acceleration factors 4, 8, and 10. More details on the image acquisition protocol are described in~\cite{wang2021recommendation}. Data of each subject comprise two different qMRI sequences: the modified-look-locker (MOLLI)~\cite{messroghli2004modified} sequence with 9 baseline images for $T_1$-mapping and the T2-prepared (T2prep)-FLASH sequence with 3 baseline images for $T_2$-mapping. The validation set contains 59 subjects, and the results are evaluated on the official platform. 

\subsection{Training Configuration}
We first train the mapping network $\mathcal{M}$, a U-Net with 256 base filters and 1 pooling layer, by Adam optimizer with an initial learning rate of $\eta_0=10^{-4}$ for 200 epochs. And then we freeze $\mathcal{M}$ for training the reconstruction backbone. We studied two different architectures for the reconstruction: the recurrent variational network (RVN)~\cite{yiasemis2022recurrent} using GRU units for regularizer and the variational network~\cite{hammernik2018learning} with U-Net as the regularizer (VN-UNet). The number of unrolled layers was set as 10 for both configurations. The loss function weighting was set as $\gamma_1=0.2, \gamma_2=0.8$. For VN-UNet, we also study the performance with relaxometry guidance, setting $\gamma_3=0.01, \gamma_4=0.1$ (VN-UNet-relax). During training, the network is trained by the Adam optimizer with an initial learning rate $\eta_0=10^{-3}$ for 400 epochs. 

Data augmentation was performed on each individual coil image as in~\cite{fabian2021data}, with random rotation $[-45^{\circ}, 45^{\circ}]$, translation $[-10\%, 10\%]$, shearing $[-20^{\circ}, 20^{\circ}]$ and vertical/horizontal flip. Additionally, we contaminate the k-space by additive Gaussian noise with a random signal-to-noise-ratio (SNR) in $[6.67, +\infty]$. The augmentation was performed with a probability of $40\%$.

\section{Results}
\subsection{Evaluation Results}
We evaluate the three aforementioned configurations: RVN, VN-UNet, and VN-UNet-relax on the validation set. For simplicity, we  only list the results on $T_1$ mapping with acceleration factor $R=4$ in Table~\ref{tab:model ablation}. The best performance was achieved by VM-UNet-relax with a PSNR of 42.59 dB. The evaluation results on all acceleration factors of both $T_1$ and $T_2$ mapping are shown in Table~\ref{tab:validation results}.
\begin{table}
    \caption{Ablation of model architecture settings on the $T_1$ mapping validation set ($R=4$). The relaxometry guided variational network with U-Nets achieved the best performance.}
    \centering
    \begin{tabular}{p{3cm} m{2.55cm}<{\centering} m{2.55cm}<{\centering} m{2.55cm}<{\centering} m{2.55cm}<{\centering}}
        \toprule
        \multicolumn{1}{l}{Model Settings}  &PSNR $(\si{dB})$ $\uparrow$ & {NMSE $(\times \si{10^{-3}})$ $\downarrow$}& {SSIM} $(\si{\%})$ $\uparrow $\\ 
        \midrule
        RVN & 41.55 & 3.4  & 97.53 \\
        VN-UNet & 42.50 & \textbf{2.6} & 97.91 \\
        VN-UNet-relax & \textbf{42.59 }& \textbf{2.6} & \textbf{97.93} \\

        \bottomrule
    \end{tabular}

    \label{tab:model ablation}
\end{table}
\begin{table}
    \caption{Evaluation results of VN-UNet-relax on the validation set. A higher acceleration factor causes a performance drop. }
    \centering
    \begin{tabular}{p{3cm} m{2.55cm}<{\centering} m{2.55cm}<{\centering} m{2.55cm}<{\centering} m{2.55cm}<{\centering}}
        \toprule
        \multicolumn{1}{l}{Dataset}  &PSNR $(\si{dB})$ & {NMSE $(\times \si{10^{-3}})$}& {SSIM} $(\si{\%})$\\ 
        \midrule
        $T_1$-$R=4$ & 43.1 & 2 & 98.1 \\
        $T_1$-$R=8$ & 37.8 & 7 & 95.6 \\
        $T_1$-$R=10$ & 36.6 & 9 & 95.1 \\
        \midrule
        $T_2$-$R=4$ & 38.9 & 3 & 97.1 \\
        $T_2$-$R=8$ & 35.2 & 6 & 94.9 \\
        $T_2$-$R=10$ & 34.4 & 7 & 94.6 \\
        \bottomrule
    \end{tabular}

    \label{tab:validation results}
\end{table}
\subsection{Qualitative Results}
\begin{figure}
    \centering
    \includegraphics{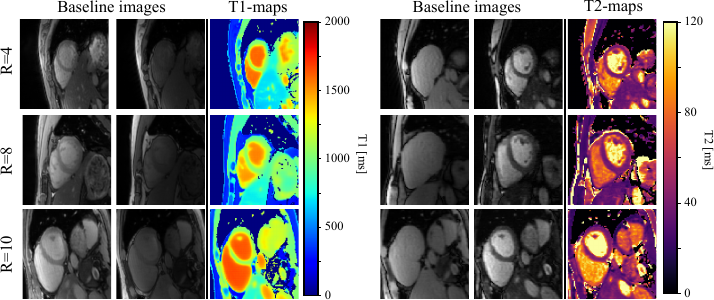}
    \caption{Qualitative results for T1 and T2 mapping sequences. The baseline images with the shortest and longest inversion/echo times are shown. The proposed method can generate both images and quantitative maps simultaneously. Perceptually, the reconstructed images of all acceleration factors are of good quality.}
    \label{fig:qualitative}
\end{figure}

We show a few exemplar reconstructed images and their corresponding quantitative maps in Fig.~\ref{fig:qualitative}. The baseline images with the shortest and longest inversion or echo times are listed. 
\section{Discussion and Conclusion}
Table~\ref{tab:model ablation} shows that the VN-UNet achieved better performance than RVN, and the performance is slightly improved by introducing relaxometry guidance. However, only image-based metrics like SSIM are provided during the validation phase. Therefore, the effect of introducing MR relaxometry-related terms in the loss function needs further investigation because the discrepancy between the predicted and the ground truth quantitative maps can not be evaluated. From Table~\ref{tab:validation results}, we observe a performance drop with the increase of acceleration factor. However, it can barely be perceived from the qualitative results shown in Fig.~\ref{fig:qualitative}.

In conclusion, we proposed a learning-based framework for qMRI reconstruction with a variational network as the reconstruction backbone and introduced an additional mapping network. The proposed framework can output both the baseline images and the mapping result simultaneously. Choosing U-Net as the regularizer achieved better performance, which is further improved by introducing MR relaxometry. 
\clearpage
%
%
\clearpage
\bibliographystyle{splncs04}
\bibliography{bibliography}
\end{document}